# A Multiwavelength View of the Nearby Calcium-strong Transient SN 2025coe in the X-Ray, Near-infrared, and Radio Wavebands

Sahana Kumar[1], Raphael Baer-Way[1,2], Aravind P. Ravi[3], Maryam Modjaz[1], Poonam Chandra[2], Stefano Valenti[3], Lindsey A. Kwok[4], Samaporn Tinyanont[5], Ryan J. Foley[6], D. Andrew Howell[7,8], Daichi Hiramatsu[9,10,11], Jennifer E. Andrews[12], K. Azalee Bostroem[13], Collin Christy[13], Noah Franz[13], Brian Hsu[13], Jeniveve Pearson[13], David J. Sand[13], Manisha Shrestha[13], Nathan Smith[13], and Bhagya Subrayan[13,14]

[1] Department of Astronomy, University of Virginia, 530 McCormick Road, Charlottesville, VA 22904, USA
[2] National Radio Astronomy Observatory, 520 Edgemont Road, Charlottesville VA 22903, USA
[3] Department of Physics and Astronomy, University of California, Davis, 1 Shields Avenue, Davis, CA 95616-5270, USA
[4] Center for Interdisciplinary Exploration and Research in Astrophysics (CIERA), 1800 Sherman Avenue, Evanston, IL 60201, USA
[5] National Astronomical Research Institute of Thailand, 260 Moo 4, Donkaew, Maerim, Chiang Mai 50180, Thailand
[6] Department of Astronomy and Astrophysics, University of California, Santa Cruz, CA 95064-1077, USA
[7] Las Cumbres Observatory, 6740 Cortona Drive, Suite 102, Goleta, CA 93117-5575, USA
[8] Department of Physics, University of California, Santa Barbara, CA 93106-9530, USA
[9] Department of Astronomy, University of Florida, 211 Bryant Space Science Center, Gainesville, FL 32611-2055, USA
[10] Center for Astrophysics|Harvard & Smithsonian, 60 Garden Street, Cambridge, MA 02138-1516, USA
[11] The NSF AI Institute for Artificial Intelligence and Fundamental Interactions, USA
[12] Gemini Observatory, 670 North A'ohoku Place, Hilo, HI 96720-2700, USA
[13] Steward Observatory, University of Arizona, 933 North Cherry Avenue, Tucson, AZ 85721-0065, USA
*Received 2026 January 26; revised 2026 May 13; accepted 2026 May 14; published 2026 June 29*

## Abstract

Calcium-strong transients (CaSTs) are a subclass of faint and rapidly evolving supernovae (SNe) that exhibit strong calcium features and notably weak oxygen features. The small but growing population of CaSTs exhibits some aspects similar to thermonuclear SNe but others similar to massive star core-collapse events, leading to intriguing questions on their physical origins. SN 2025coe is one of the nearest CaSTs discovered to date, and our coordinated multiwavelength observations obtained days to weeks postexplosion reveal new insights into these enigmatic transients. With the most robust near-IR (NIR) spectroscopic time series of a CaST collected to date, SN 2025coe shows spectral signatures characteristic of Type Ib SNe (SNe Ib; i.e., He-rich stripped-envelope SNe (SESNe)). SN 2025coe is the third X-ray-detected CaST and our analysis of Neil Gehrels Swift Observatory X-ray data suggests interaction with $0.12 \pm 0.11\, M_\odot$ of circumstellar material (CSM) extending to at least $2 \times 10^{15}$ cm ($\sim$30,000 $R_\odot$), while our analysis of the 1–240 GHz radio nondetections gives an outer radius of that CSM of at most $\sim 5 \times 10^{15}$ cm. This inferred nearby high-density CSM extending out to $(3.5 \pm 1.5) \times 10^{15}$ cm is similar to that seen in the other two X-ray-detected CaSTs, and its presence suggests that either intensive mass loss from a massive star or some exotic pre-SN mass ejection may be a common feature of this subclass. Our work also expands upon recent studies of the optical properties of SN 2025coe and explores our current understanding of different progenitor systems that could possibly produce CaSTs.

## 1. Introduction

Ca-strong transients (CaSTs) are a growing subclass of faint supernovae (SNe) with puzzling origins. Compared to other types of SNe, CaSTs are quite rare and their spectra exhibit prominent [Ca II] emission features at both photospheric and nebular phases (A. V. Filippenko 1997; M. Modjaz et al. 2019; K. De et al. 2020). Over the past 2 decades, ∼40 SNe have been classified as CaSTs based on optical spectra taken weeks postexplosion (H. B. Perets et al. 2010; M. M. Kasliwal et al. 2012) and current estimates predict CaSTs occur at a rate of $\sim$2 decade$^{-1}$ within a distance of ∼25 Mpc (C. Frohmaier et al. 2018).

Several CaSTs have been identified in the literature as Ca-rich transients (CaRTs) due to their prominent [Ca II] emission lines. Based on previously published works (K. J. Shen et al. 2019), we opt to use "Ca strong" as opposed to "Ca rich" due to the easily excitable nature of Ca and the fact that the [Ca II] emission features in CaSTs are strong when compared to [O I]. Therefore, CaSTs are spectroscopically characterized by weak O lines in addition to prominent Ca lines. Estimates of the calcium abundances in these objects do not reveal an excess of calcium, but rather a dearth of oxygen, confirming the CaST designation rather than CaRT (W. V. Jacobson-Galán et al. 2020a).

The progenitors of CaSTs are an active area of debate because their observed properties do not neatly fit into the categories of core-collapse or thermonuclear SNe. For example, the majority of CaSTs are found in old stellar populations in the outskirts of elliptical host galaxies (H. B. Perets et al. 2011; R. J. Foley 2015; R. Lunnan et al. 2017), however a growing number of CaSTs has been discovered in spiral, star-forming hosts (K. De et al. 2020). Furthermore, some CaSTs have double-peaked light curves (W. V. Jacobson-Galán et al. 2020a, 2022; A. P. Ravi et al. 2026) which may indicate interaction with nearby circumstellar material (CSM) or a shock-heated envelope that is expanding and cooling (V. Morozova et al. 2017; C. Pellegrino et al. 2023; A. Crawford et al. 2025). Previously published studies suggest some CaSTs with evidence of CSM interaction may be the result of the core-collapse of a low-mass massive star (D. Milisavljevic et al. 2017; S. J. Prentice et al. 2020), whereas other studies posit that CaSTs with large offsets from their host galaxies may be of thermonuclear origins (W. V. Jacobson-Galán et al. 2020a). Despite the growing population of known CaSTs, it is currently unclear whether these transients are the result of the core-collapse of a massive star or thermonuclear explosion of a white dwarf (WD).

[14] LSSTC Catalyst Fellow.

At peak brightness, the optical spectra of CaSTs often resemble stripped-envelope SNe (SESNe; i.e., SNe of Types IIb, Ib, and Ic), with some papers introducing terminology such as "Ca-IIb" (K. K. Das et al. 2023). Other optical characteristics of CaSTs include peak magnitudes of $-14$ to $-16.5$ mag and rapid rise times of $t_r < 15$ days (S. Taubenberger 2017; W. V. Jacobson-Galán et al. 2020a). Thus, CaSTs exhibit intrinsically fainter peak magnitudes than other types of SNe and have quickly declining optical light curves, probably due to lower ejecta and $^{56}$Ni masses as inferred by, e.g., K. De et al. (2021). These observations are difficult to reconcile with a single, massive star progenitor, pointing instead toward a massive WD progenitor or highly stripped massive star in a binary system (K. J. Shen et al. 2019).

The two closest CaSTs have shown signs of interaction with CSM that was lost before the explosion (W. V. Jacobson-Galán et al. 2020a, 2022). When SN ejecta run into a dense CSM, forward and reverse shocks are generated that power radio synchrotron emission and thermal free–free and nonthermal X-ray emission (R. A. Chevalier 1998). Radio and X-ray observations give direct information on the CSM density and corresponding mass-loss rate of the progenitor (in a more precise manner than optical estimates as the radio and X-ray emission comes from interaction alone, in the absence of a central engine), providing direct constraints on progenitors as different mass-loss rates can be used to rule out or allow certain progenitor types (see, e.g., P. Chandra 2025).

Near-infrared (NIR) data may be particularly useful for studying CaSTs because they can directly probe the presence of helium in the outer layers of the exploding star, as the 2.058 $\mu$m He line is not blended with other features and one of the strongest lines (e.g., L. Dessart et al. 2012; S. Hachinger et al. 2012; M. Williamson et al. 2021). Comparing the strengths and velocities of helium lines can allow for insight into differences in opacity, which may reflect changing physical conditions within the ejecta (J. C. Wheeler et al. 1998) that are otherwise difficult to understand in rapidly evolving objects like CaSTs. To date, only three CaSTs have previously published NIR spectra: iPTF15eqv (D. Milisavljevic et al. 2017), SN 2019ehk (W. V. Jacobson-Galán et al. 2020a), and SN 2021gno (W. V. Jacobson-Galán et al. 2022) and two of these have X-ray detections (W. V. Jacobson-Galán et al. 2020a, 2022). All three NIR-observed objects exhibit prominent NIR He lines in addition to the characteristic Ca lines.

In this work, we present the most robust NIR spectral time series of a CaST to date along with multiepoch X-ray and radio follow-up. SN 2025coe is located in the outskirts of its host galaxy, offset from the host center by $\sim$40 kpc. Discovered on 2025 February 24 within hours of explosion and initially classified as an SN Ib, SN 2025coe was later reclassified as a CaST associated with NGC 3277 (M. Andrews et al. 2025a). Located at a distance of 25 $\pm$ 9.3 Mpc using a Tully–Fisher estimate (H. W. Edler et al. 2024), SN 2025coe is also the third-nearest CaST discovered to date.

The paper is laid out as follows: X-ray, NIR, and radio observations are presented in Section 2. X-ray analysis is presented in Section 3.1. Identification of NIR spectral features are presented in Section 3.2, followed by a comparison to NIR spectral templates in Section 3.3 using synthetic spectra of potential origin scenarios for CaSTs. The presence of CSM interaction is discussed in Sections 3.1 and 3.4 using X-ray and radio observations at early and late times. Finally, in Section 3.5, we combine observational clues across wavelength regimes to construct a possible physical explanation of SN 2025coe.

## 2. Observations

Once SN 2025coe was identified as a nearby CaST, we coordinated follow-up observations across multiple wavelength regimes. In the following sections, we present X-ray, NIR, and radio observations of SN 2025coe. Complementary optical photometry and spectroscopy is published in an accompanying paper (A. P. Ravi et al. 2026), and is the source of all optical properties of SN 2025coe used in this study.

### 2.1. X-Rays

SN 2025coe was observed at X-ray wavelengths using the X-Ray Telescope (XRT) on board the Neil Gehrels Swift Observatory (Swift) spacecraft. Observations were taken at 11 separate postexplosion epochs covering 2–100 days, but some of these observations were quite short (<1 ks) and closely spaced in time. We thus opt to combine observations where $\delta t < 2$ days or $\delta t/t < 0.2$ and end up with five distinct epochs of observations with at least 1.5 ks of exposure time. We reduced the data using standard procedures with `xrtpipeline`.[15]

Although there are no preexplosion X-ray images of the SN site, SN 2025coe is very isolated with no significant X-ray sources within $\sim$100″. SN 2025coe is detected in the X-ray images at 2.98 and 8.16 days postexplosion with a signal-to-noise ratio > 2.5 (all measurements assume Poisson statistics to estimate the background). Figure 1 shows a $10' \times 10'$ view of the field showing the SN site (for detections and nondetections) that is offset by $\sim$316″ from the X-ray bright center of the host galaxy, NGC 3277. Independent analysis of near-UV and far-UV observations by A. P. Ravi et al. (2026) also showed no evidence for strong UV flux near the SN location, further emphasizing the remote location of SN 2025coe and reducing the likelihood of host galaxy contamination.

We use SOSTA as part of HEASoft[16] to find detection rates and 3$\sigma$ upper limits at all epochs while accounting for vignetting and point-spread functionlosses. We use a 25″ region centered around the SN location for source detection, with the

[15] swift.gsfc.nasa.gov/analysis/xrt_swguide_{v}{1}_2.pdf
[16] https://heasarc.gsfc.nasa.gov/docs/software/lheasoft/

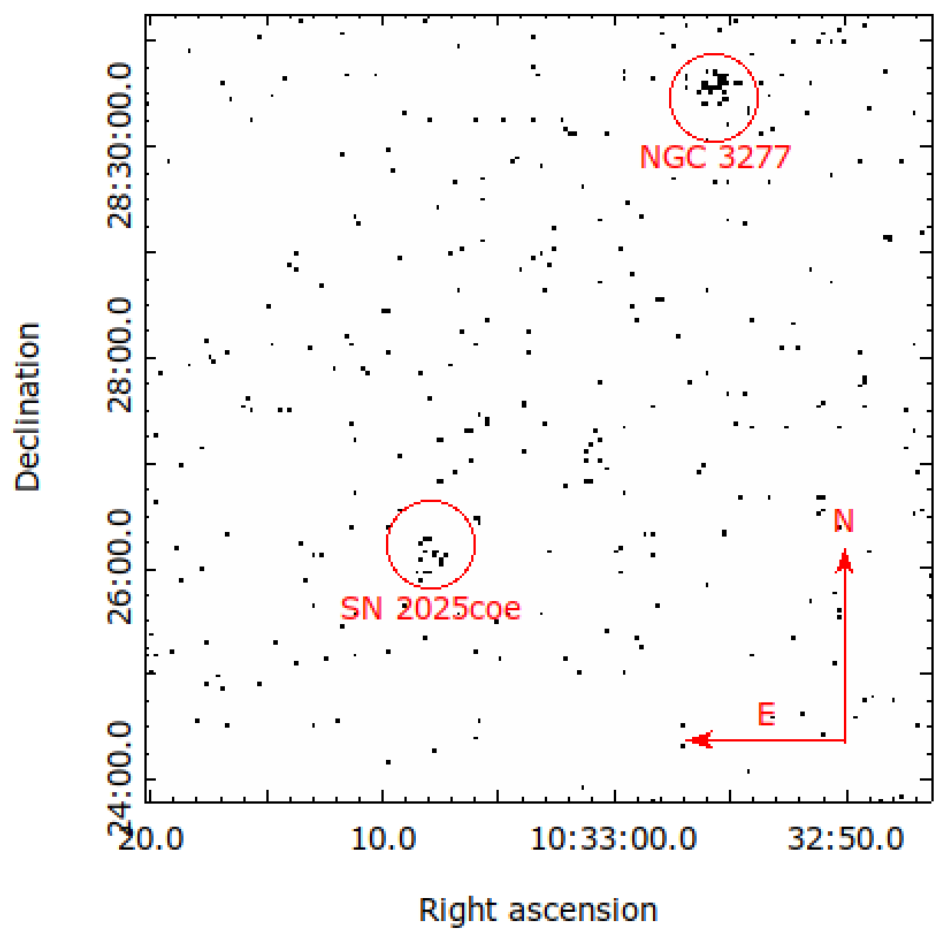

A merged X-ray image from the SN 2025coe detections from 1-12 days post-explosion.

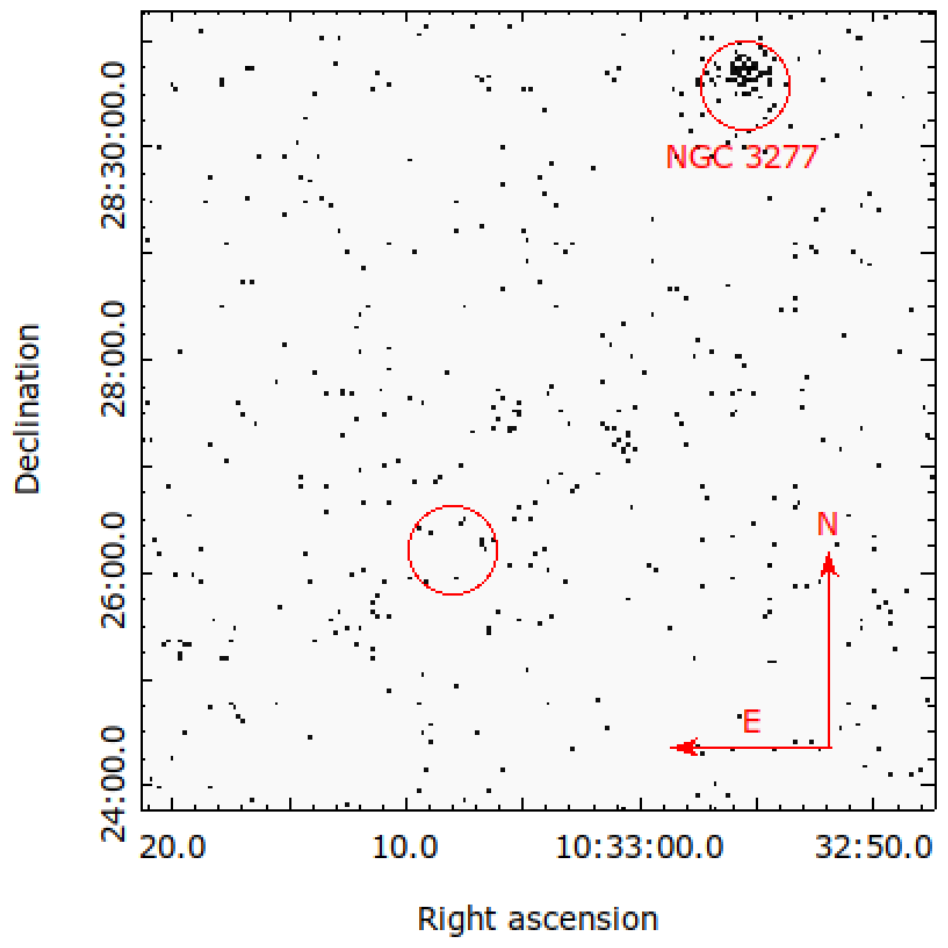

A merged X-ray image from the later epochs (20-92) days of SN 2025coe, when the supernova was not detected.

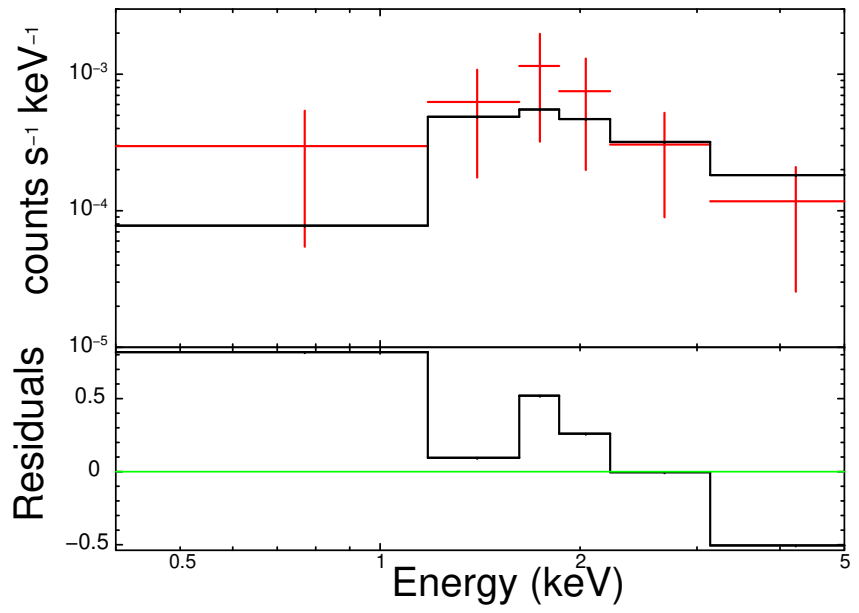

X-ray spectrum from detections and best-fit model background estimated from an annulus between 50″ and 100″. Both detections yield a count rate $\sim 2.4 \times 10^{-3}$ counts $s^{-1}$. Given that both of these early observations are the only observations with detections, we opt to combine these to obtain enough counts to fit a spectral model. We note that the spectrum obtained from individual observations does not differ significantly from the parameters we obtain for the combined spectrum; it only has larger error bars.

**Figure 1.** A view of the merged Swift X-ray images showing detections and nondetections of SN 2025coe, including the location of the host galaxy center and the 25″ regions used for the source extraction. Based on the postdetection image, there is no detectable X-ray source near the SN that could potentially contribute to the observed flux. The bottom panel shows the absorbed power-law fit to the combined X-ray spectrum of SN 2025coe by combining observed data from 2.98 and 8.17 days postexplosion for the spectral fitting.

Given the best-fit power law to SN 2019ehk (W. V. Jacobson-Galán et al. 2020a), we fit a power law with the Swift data products tool (P. A. Evans et al. 2009) at the location of the SN and find that there is a significant absorbing column density from the SN environment of $(1.5^{+1.5}_{-3.8}) \times 10^{22}$ $cm^{-2}$. This value is much larger than the Galactic column density at the coordinates of SN 2025coe of $2.2 \times 10^{20}$ $cm^{-2}$ using the method of R. Willingale et al. (2013). We note that fits for thermal emission give a similar flux with a slightly worse $\chi^2$. We find a best-fit power-law index $\Gamma = 2.0^{+2.6}_{-1.7}$. We use this power-law index to convert each count rate to a luminosity, given the large error bars and limited counts in general. We also use the best-fit $\Gamma/N_H$ to convert our upper limits on the count rate (also found using the same regions described above) at later times to upper limits on flux. Results are reported in Table 1 and shown in Figure 2. We show the absorbed power-law fit in Figure 1. We interpret these results in Section 3.1.

**Table 1**
Swift/XRT Observations of SN 2025coe

| Days Postexplosion (days) | Count Rate (counts $s^{-1}$) | Exp. Time (s) | Unabsorbed Flux (erg $cm^{-2}$ $s^{-1}$) |
|---|---|---|---|
| 2.98 ± 1.16 | $(2.38 \pm 1.1) \times 10^{-3}$ | 3297 | $(2.84 \pm 1.2) \times 10^{-13}$ |
| 8.17 ± 2.00 | $(2.48 \pm 1.31) \times 10^{-3}$ | 3849 | $(2.96 \pm 1.49) \times 10^{-13}$ |
| 20.18 ± 0 | $<6.96 \times 10^{-3}$ | 1880 | $<1.08 \times 10^{-12}$ |
| 30.78 ± 4.74 | $<3.76 \times 10^{-3}$ | 2700 | $<5.82 \times 10^{-13}$ |
| 92.70 ± 0 | $<3.48 \times 10^{-3}$ | 3149 | $<5.39 \times 10^{-13}$ |

### 2.2. Near-infrared Spectra

The first two NIR spectroscopic observations of SN 2025coe were obtained by the Keck Infrared Transient Survey (S. Tinyanont et al. 2024) using Keck II + the NIR Echelette Spectrometer (NIRES; I. S. McLean et al. 1998). These observations were obtained on 2025 March 7 and March 20, corresponding to approximately 10 and 23 days postexplosion, respectively. Further information on the observations and NIRES data reduction can be found in S. Tinyanont et al. (2024). An additional NIR spectrum of SN 2025coe was taken on 2025 April 10 with the MMT and Magellan Infrared Spectrograph (MMIRS; B. McLeod et al. 2012). These observations were reduced using the MMIRS pipeline (I. Chilingarian et al. 2015), and the resulting one-dimensional spectra were telluric and absolute flux corrected using the method described (W. D. Vacca et al. 2003) with the `XTELLCOR_` tool (part of the `Spextool` package; M. C. Cushing et al. 2004) using a standard AV0 star observed at a similar air mass and time. We were not able to recover data in the telluric region as has been done for other SNe (S. Kumar et al. 2026). These NIR spectra are presented in Figure 3 and Table 2.

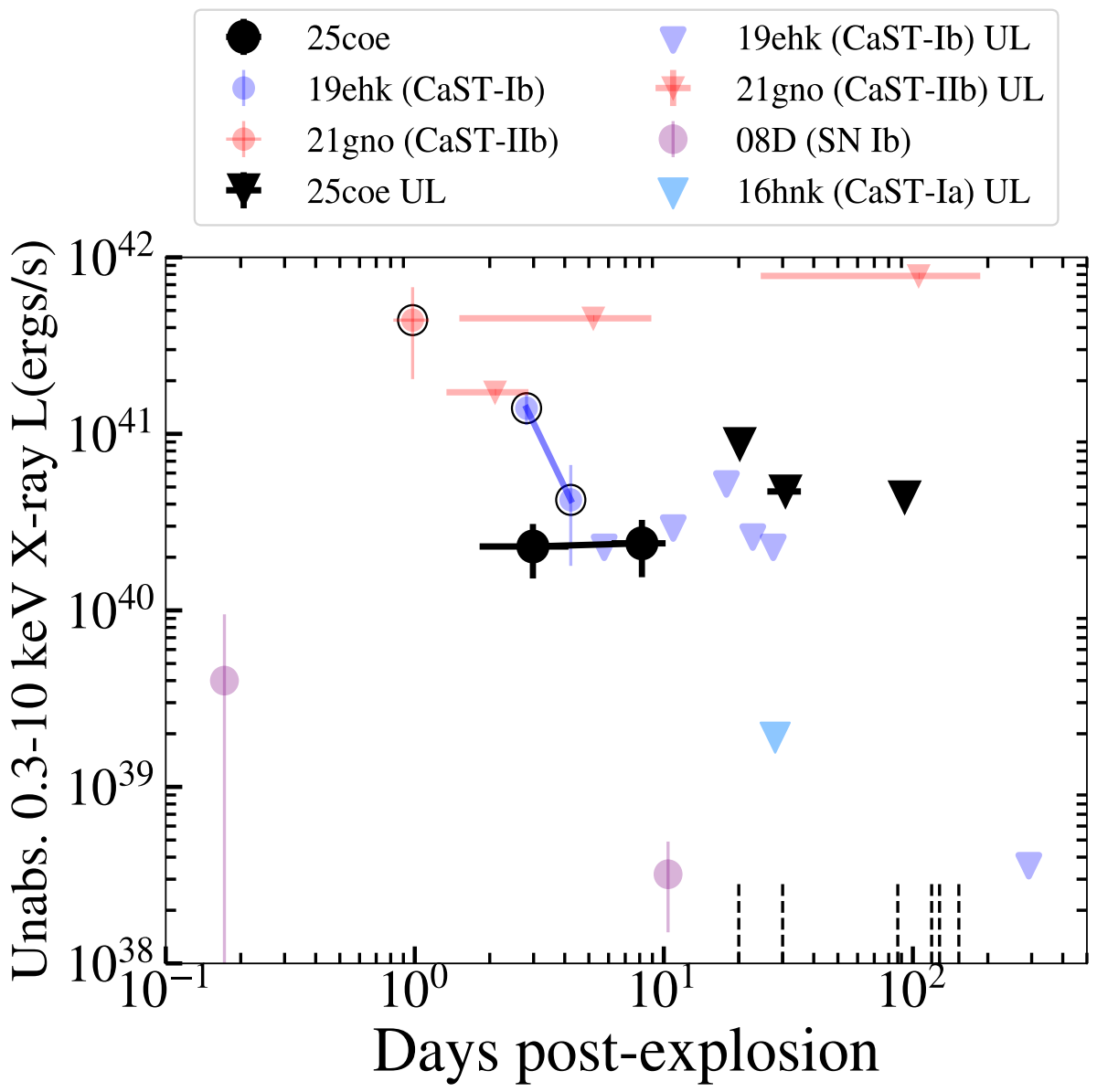

**Figure 2.** The X-ray detections of SN 2025coe in context with the other two X-ray-detected CaSTs SN 2019ehk and SN 2021gno (W. V. Jacobson-Galán et al. 2020a, 2022). We also show the nondetection of the CaST-Ia SN 2016hnk (P. H. Sell et al. 2018; W. V. Jacobson-Galán et al. 2020b) and the X-ray evolution of the SN Ib SN 2008D. The vertical dashed lines denote the epochs of radio nondetections for SN 2025coe, and triangular markers denote the upper limits determined by the observed X-ray nondetections.

### 2.3. Radio

We monitored SN 2025coe across the radio spectrum with various telescopes for 3 months postexplosion. The first observations were taken at 20 days postexplosion with the Submillimeter Array (SMA) at 240 GHz, with data reduced using standard techniques from the Common Astronomy Software Applications package. We observed with the Giant Meterwave Radio Telescope (GMRT) at 1.25 GHz at 30 and 90 days postexplosion, and with the Very Large Array (VLA) with the *S*, *C*, and *X* bands (3–10 GHz) at later epochs at >100 days postexplosion. All data were reduced in the usual manner for their respective telescopes, with the standard pipelines used for GMRT and VLA data (B. R. Kent et al. 2020; R. Kale & C. H. Ishwara-Chandra 2021). Given the lack of bright radio sources in the field, no self-calibration was done for the VLA data and a few rounds of phase-only self-calibrations were performed for the GMRT data. All observations yielded nondetections. The nondetections are reported in Table 3. All nondetections are $3\sigma$ (taken from $3\times$ the rms in a region multiple times the size of the beam around the SN location). We show the nondetections in context with other CaSTs in Figure 4.

### 2.4. Optical Parameters

Our companion paper (A. P. Ravi et al. 2026), presents a comprehensive analysis of the optical dataset of SN 2025coe. We summarize their conclusions here, which will be used to provide additional context and observational information while discussing the multiwavelength behavior of SN 2025coe.

1. Multiband photometry of SN 2025coe reveals the presence of two peaks, making it the sixth double-peaked CaST.
2. SN 2025coe is significantly offset from the likely host galaxy NGC 3277, making a massive star origin more difficult to explain due to the lack of star formation in the area. However, there are nearby fainter sources that could be satellites of NGC 3277 and the host galaxy of SN 2025coe itself.
3. Modeling the bolometric light curve, they find that the first peak (at ∼2 days after explosion) can be explained by the shock cooling of a compact envelope with a radius of ∼6–40 $R_\odot$, and a mass of ∼0.1–0.2 $M_\odot$, and/or close-in CSM interaction ($R_{\rm CSM} \lesssim 6 \times 10^{14}$ cm). The second peak (at ∼11 days after explosion) is powered by radioactive decay from an ejecta mass of ∼0.4–0.5 $M_\odot$ and a $^{56}$Ni mass of ∼0.014 $M_\odot$.
4. Like other members of this class, SN 2025coe rapidly evolves from the photospheric phase dominated by He I P Cygni profiles to nebular phase spectra dominated by [Ca II] $\lambda\lambda$7291, 7323 and weak [O I] $\lambda\lambda$6300, 6364. This evolution occurs within ∼10 days, which is very rapid compared to other types of SNe.
5. Simultaneous line profile modeling of [Ca II] $\lambda\lambda$7291, 7323 and [O I] $\lambda\lambda$6300, 6364 at nebular phases show that an asymmetric core-collapse explosion of a low-mass He-core progenitor ($\lesssim$3.3 $M_\odot$) can explain the observed line profile shapes, indicating the possibility of a low-mass massive star origin. Alternatively, lack of local star formation at the site of the SN explosion combined with a low ejecta mass is also consistent with a thermonuclear explosion due to a low-mass hybrid HeCO WD + CO WD merger.

In general, these results are consistent with the interpretations from our dataset, which we describe below.

## 3. Discussion

Despite the growing number of CaSTs discovered over the past few decades (K. De et al. 2020), much remains unknown about their observed properties, particularly in the NIR, X-ray, and radio. One challenge is that CaSTs are often classified as young SESNe shortly after discovery, but the characteristic strong Ca and weak O spectral features do not emerge until days to weeks later. For example, SN 2025coe was initially classified as an SN Ib (M. Andrews et al. 2025b) whereas SN 2021gno was initially classified as an SN IIb (T. Hung et al. 2021).

The rapid decline of the optical light curves is another identifying feature of CaSTs, but this decline occurs too late to trigger multiwavelength follow-up observations prior to peak brightness. Examining the time evolution across multiple wavelength regimes provides new insights on the origins of these transients and may yield clues on how to identify CaSTs at earlier phases.

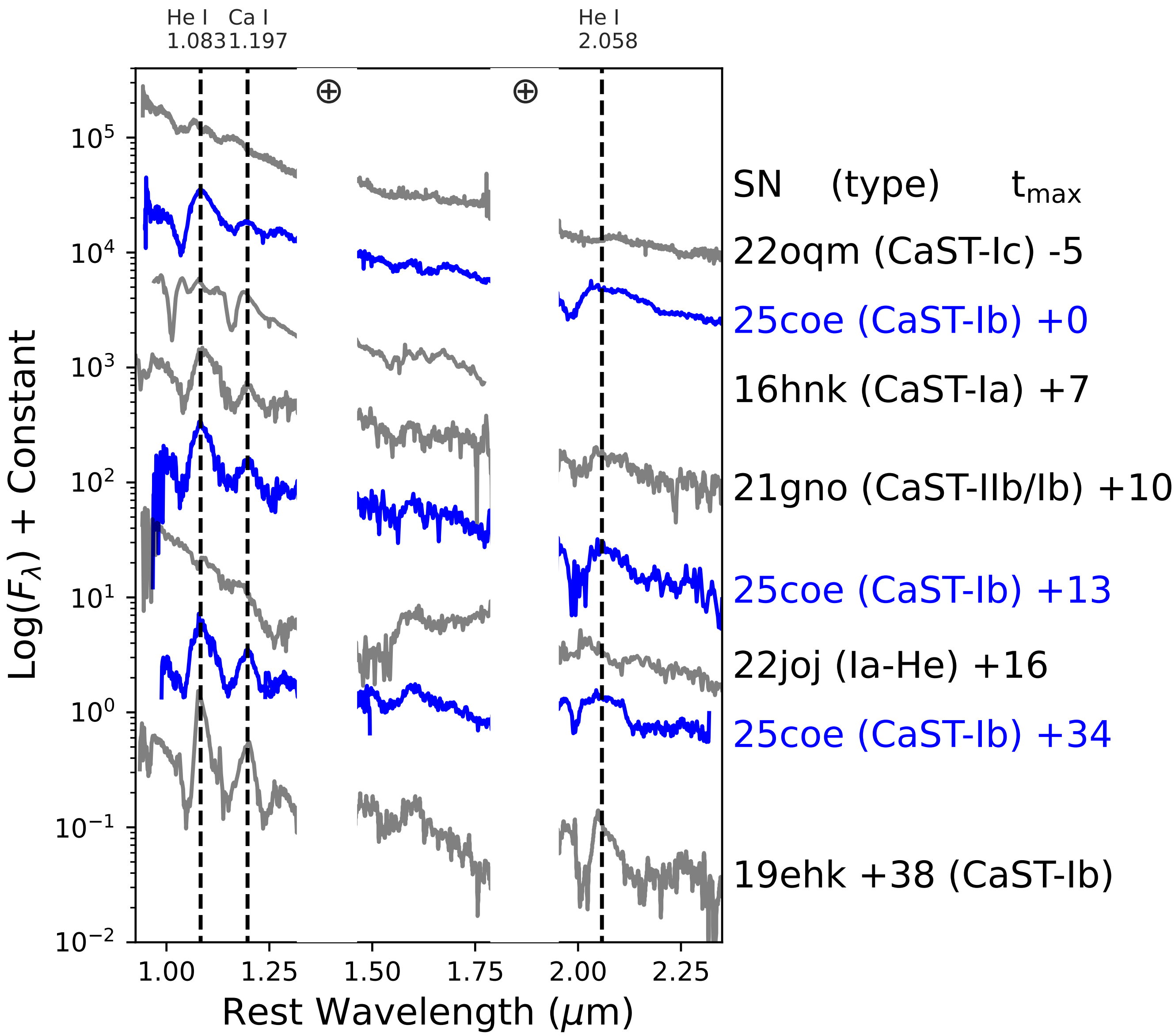

**Figure 3.** All NIR spectra of SN 2025coe compared to other CaSTs. To date, SN 2025coe has the most robust NIR spectral times series of any CaST. All phases are listed with respect to optical maximum of the secondary, nickel-powered light-curve peak. At earlier times, SN 2025coe does not show many similarities to CaRTs Ca-Ic 2022oqm (S. K. Yadavalli et al. 2024) or Ca-Ia 2016hnk (L. Galbany et al. 2019), but later evolves to resemble CaSTs Ca-IIb 2021gno (W. V. Jacobson-Galán et al. 2022; K. Ertini et al. 2023; A. Crawford et al. 2025) and Ca-Ib 2019ehk (W. V. Jacobson-Galán et al. 2020a). SN 2022joj is a Type Ia SN (SN Ia) that has been identified as a potential helium shell detonation candidate (E. Padilla Gonzalez et al. 2024), and is included here for comparison.

(The data used to create this figure are available in the online article.)

**Table 2**
Near-infrared Spectroscopy of SN 2025coe

| UT Date (YYYY-MM-DD) | Phase (days) | Days Postexplosion (days) | Exp. Time (s) | Telescope + Instrument |
|---|---|---|---|---|
| 2025-03-07 | +0 | 10 | 800 | Keck + NIRES |
| 2025-03-20 | +13 | 23 | 2880 | Keck + NIRES |
| 2025-04-10 | +34 | 45 | 2400 | MMT + MMIRS |

**Note.** The phases listed here are defined relative to the *r*-band maximum from the optical light curves and explosion date published in A. P. Ravi et al. (2026). This phase convention is chosen for the NIR spectra in order to provide useful comparison to previously published optical and NIR spectra. The spectral evolution with respect to peak brightness is useful when considering the composition of the ejecta and how CaSTs evolve compared to other types of SNe.

### *3.1. X-Ray Modeling*

SN 2025coe is the third CaST ever detected at X-ray wavelengths, with these three objects constituting the three nearest CaSTs ever discovered. The fact that X-ray emission was seen in all three objects suggests that X-ray emission is a common feature of this class as shown in Figure 2, but a larger

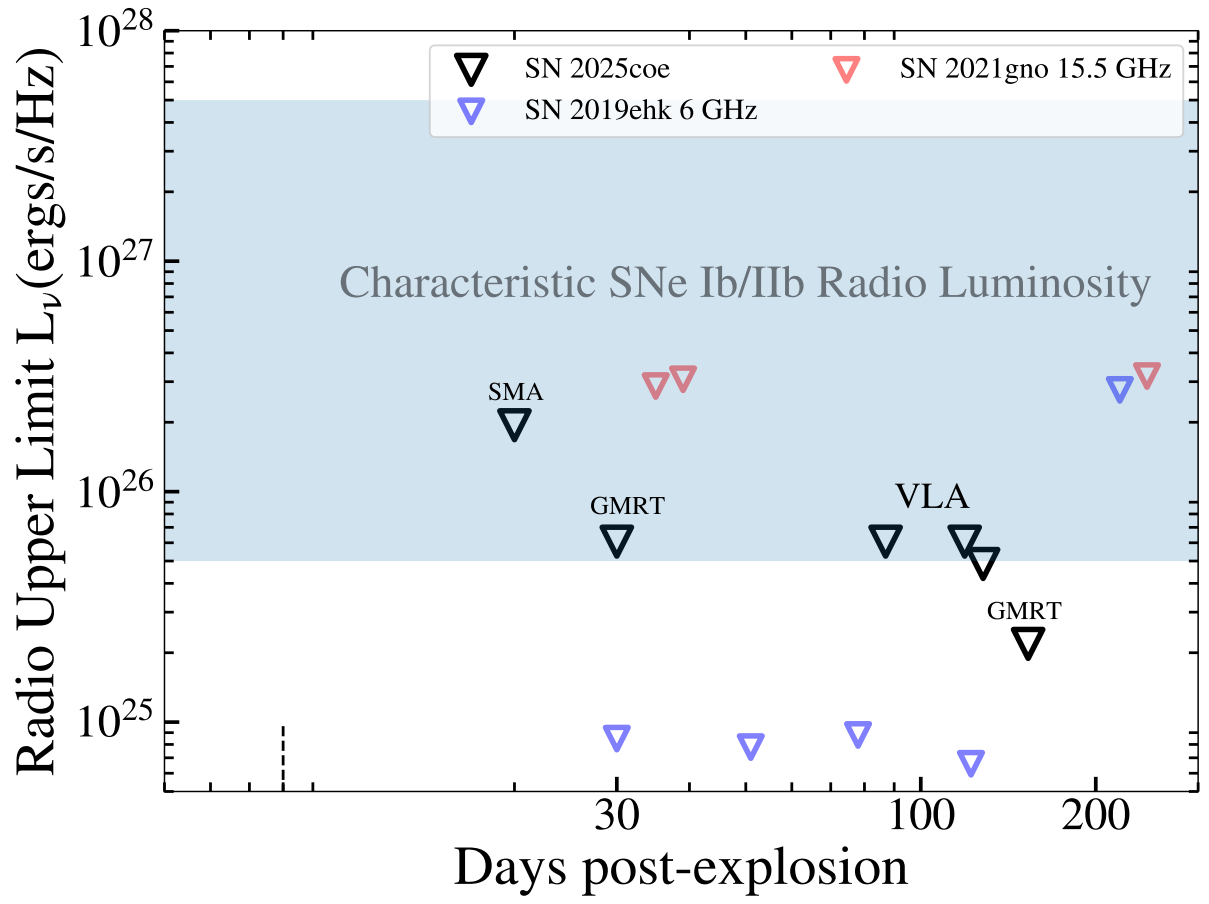

**Figure 4.** A view of the upper limits on radio luminosity for three CaSTs: SN 2019ehk, SN 2021gno, and SN 2025coe. The SN 2019ehk and SN 2021gno data are from W. V. Jacobson-Galán et al. (2020a, 2022). To date, no CaST has ever been detected at radio wavelengths. The VLA data are at *S*, *C*, and *X* band (3–10 GHz), the GMRT data are at 1.25 GHz, and the SMA data are at 240 GHz. The dashed line denotes the last X-ray detection of SN 2025coe. We show the typical radio luminosity space of SNe Ib and IIb as well (M. F. Bietenholz et al. 2021). The radio luminosity constraints are quite deep, and an order of magnitude lower than the typical luminosity of a detected SESN. These nondetections clearly suggest that SN 2025coe was not interacting with dense CSM at late times, similar to SN 2019ehk and SN 2021gno (W. V. Jacobson-Galán et al. 2020a, 2022).

**Table 3**
Radio Observations of SN 2025coe

| UT Date | Epoch (days) | Frequency (GHz) | $F_\nu$ (mJy) | Image rms ($\mu$Jy) | Telescope |
|---|---|---|---|---|---|
| 2025 Mar 16 | 20 | 240 | <0.1 | 20 | SMA |
| 2025 Mar 26 | 30 | 1.25 | <0.075 | 20 | uGMRT |
| 2025 May 22 | 87 | 1.25 | <0.075 | 22 | uGMRT |
| 2025 Jun 19 | 119 | 15 | <0.060 | 20 | VLA |
| 2025 Jun 28 | 128 | 10 | <0.027 | 10 | VLA |
| 2025 Jul 28 | 153 | 6 | <0.015 | 13 | VLA |

**Note.** All upper limits are $3\sigma$. Epochs are defined relative to the date of explosion defined by A. P. Ravi et al. (2026).

sample of nearby objects is needed. All three objects were detected with Swift/XRT at early times, but their X-ray light curves are distinct. For SN 2025coe, the X-ray detections shown in Figure 2 at ∼3 and 8 days postexplosion are at a similar unabsorbed flux level.

The observed X-ray plateau could be due to a variety of reasons: decreasing absorption, the X-ray emission peaking between 3 and 8 days, decreasing temperature of the X-ray emission causing more of the luminosity to fall in the Swift/XRT 0.2–10 KeV range, a constant-density CSM ($\rho_{\rm CSM} \sim \propto r^0$), and/or a radiative shock (R. A. Chevalier 1998). Without multiple epochs of detections beyond these first two epochs, it is not possible to infer the density profile of the CSM from this early plateau. What is notable is that the best-fit power-law photon index $\Gamma \sim 2$, suggesting a slightly softer X-ray-emitting component with thermal temperature ∼ 5 keV, contrary to what was seen in SN 2019ehk and SN 2021gno, which had thermal temperatures > 10 keV (W. V. Jacobson-Galán et al. 2020a). The implied lower temperature in SN 2025coe implies that the emission is likely coming from a reverse shock (P. Chandra 2025), but the large error bars prevent definitive confirmation.

**Table 4**
Swift/XRT Bremsstrahlung Modeling Details of SN 2025coe

| $N_{\rm H}$ ($10^{22}$ cm$^{-2}$) | Temperature (keV) | Normalization | CSM Density (g cm$^{-3}$) |
|---|---|---|---|
| 1.5 (frozen) | 5 (frozen) | 0.045 ± 0.026 | (2.0 ± 1.9) × $10^{-14}$ |
| 1.5 (frozen) | 20 (frozen) | 0.034 ± 0.021 | (1.7 ± 1.7) × $10^{-14}$ |
| 1.5 (frozen) | 50 (frozen) | 0.037 ± 0.024 | (1.8 ± 1.8) × $10^{-14}$ |

**Note.** We select the 5 keV model for parameter estimation as this is the best-fit model based on $\chi^2$.

We can use a bremsstrahlung approximation to estimate the total CSM mass that produced this X-ray emission and compare that total mass with the CSM inferred for other X-ray-luminous CaSTs. We approximate a nonthermal power law with a photon index of two by using a thermal bremsstrahlung model assuming a temperature of 5 keV. Fitting a bremsstrahlung model to the data with Xspec, we obtain an estimate of the emission measure and use this to find an estimate of the CSM mass and density using Equation (2) from D. Brethauer et al. (2022).

Given that the temperature and column density are not well constrained, we conduct fits at three different temperatures while keeping the column density fixed to the best-fit power-law value to evaluate the impact of the uncertain temperature. We attempt fits with $T = 5$, 20, and 50 keV (where the 20 and 50 keV models would correspond to a forward shock emission scenario). We find the resulting CSM density and emission measure estimates agree within ±20%, and thus suggest that our derived density measurement is consistent (with the large error bars) regardless of the true temperature of the X-ray-emitting material. Table 4 lists the model parameters and results for all three temperatures. The 5 keV model is adopted as the final best-fit model as it provides the best fit in terms of $\chi^2$ to the combined X-ray spectrum.

For the volume of the CSM, we assume a spherical CSM geometry such that $V = (4/3)\pi R^2 \Delta R$, while also assuming $\Delta R = R$. We note that the CSM may in fact be aspherical given the lack of optical signatures of interaction (A. P. Ravi et al. 2026), and if we introduce a filling factor $f$ to account for clumping such that $V_{\rm eff} = f V_{\rm spherical}$ (R. A. Chevalier 1998) the derived CSM density of the clumps would be higher by a factor 2–10 for reasonable $f = 0.1$–$0.5$. In any case, for this radius of the CSM, we take the midpoint between the detections of 5.57 days, and assume a shock speed of $\sim$30,000 ± 10,000 km s$^{-1}$ based on what is typically seen at early times in interacting SESNe (R. A. Chevalier & C. Fransson 2017) and assumed in another CaST-Ib (W. V. Jacobson-Galán et al. 2020a). We find $R = 1.44 \times 10^{15}$ cm, but note that the outer extent of the CSM based on the detection at 8 days would be $2.1 \times 10^{15}$ cm.

We emphasize that our inferred CSM radii are directly dependent of our assumption of the shock speed, and thus could be lower and more consistent (for example, a 15,000 km s$^{-1}$ shock speed) with the blackbody radii estimated by A. P. Ravi et al. (2026) using UV + optical spectral energy distribution fitting. From the bremsstrahlung fit with Xspec (obtaining errors from $\Delta\chi^2$), we find an emission measure of $1.21 \pm 0.7 \times 10^{66}$ cm$^{-3}$. Without any constraint on the CSM composition due to the lack of any optical spectral signatures of CSM, we assume solar abundances and fully ionized CSM ($\mu_e = 1.25$ and $\mu_I = 1.15$) for simplicity. Using these assumptions and the assumed $R = 1.4 \times 10^{15}$ cm (incorporating errors from the fact that the fit was done for combined epochs from 1.82 to 10.2 days), we find $M_{\rm CSM} = 0.12 \pm 0.11\,M_\odot$ from $\rho_{\rm CSM} = (1.96 \pm 1.90) \times 10^{-14}$ g cm$^{-3}$. Assuming a helium-rich CSM would increase these numbers by a factor ∼2. In either case, the CSM mass derived would imply a mass-loss rate of ∼0.2–0.5 $M_\odot$ yr$^{-1}$ if the mass was lost directly in the last 100–200 days before the explosion given the fact that it was only detected up to ∼10 days after the explosion, assuming an ejecta speed/CSM speed ratio of 10–20 (CSM speed of ∼500–1500 km s$^{-1}$), as has been seen for other interacting objects (P. Chandra 2025).

In other words, to create a high-density CSM that only extends to $\sim10^{15}$ cm in a core-collapse scenario, the progenitor of SN 2025coe must have lost mass at a very enhanced rate in the last months to years prior to the explosion. Without an accurate estimate of the CSM speed, this time frame cannot be constrained more precisely.

After these initial X-ray detections, SN 2025coe was not detected again. The nondetections starting from 20 to 92 days postexplosion are at higher fluxes than the initial detections due to the shorter exposure times with Swift, and thus do not allow for strong constraints on the late-time mass-loss rate. However, the nondetections line up with the time frame of nondetections at radio wavelengths (with the first radio epoch at 20 days postexplosion, see Section 3.4), suggesting that the CSM interaction has likely ended and that the dense CSM was relatively confined near to the progenitor system. The lack of extended CSM aligns with what has been seen for other CaSTs, although we emphasize again that the X-ray detection as late as 8 days is the latest ever detection for a CaST and suggests the presence of CSM out to larger radii than in SN 2021gno and SN 2019ehk (although SN 2019ek had a CSM radius of $\sim10^{15}$ cm as well, and is discussed further in Section 3.5.

Furthermore, SN 2025coe seems to have ∼20 times more CSM mass (although with large error bars) than SN 2019ehk and SN 2021gno. It is important to note that we assume an essentially constant-density CSM without any constraints on the density profile, so the larger CSM radius naturally implies a larger CSM mass. However, the total CSM mass is still relatively low at ∼0.1 $M_\odot$ and does not exceed the progenitor mass expected in either the core-collapse or thermonuclear scenario. The 0.1 $M_\odot$ mass is still compatible with the amount of material that could surround a thermonuclear explosion with pollution from a helium nova (C.-G. Touchard-Paxton et al. 2025), or the amount that could be ejected in late-stage mass loss from a massive star. The fact that there is nearby high-density CSM, however, is very different from most SESNe, which show mass-loss rates of $\sim10^{-4}$–$10^{-6}$ $M_\odot$ yr$^{-1}$, whereas our measurements imply a mass-loss rate $> 10^{-1}$ $M_\odot$ yr$^{-1}$. In order to explain why the dense and nearby CSM in SN 2025coe did not produce any photoionized narrow emission lines in the optical spectra (A. P. Ravi et al. 2026), we speculate that the X-ray-emitting CSM is in fact either distributed highly asymmetrically or that there is significant clumping of the CSM.

### 3.2. Near-infrared Line Identification and Fitting

As described, we obtained three NIR spectra of SN 2025coe. The main features of interest are the He I 1.083 and 2.058 μm features, seen in other CaSTs and SNe Ib and IIb. We show the spectra in Figure 3 along with other NIR spectra of CaSTs. Given that the emission peak lines up nearly exactly with zero velocity for the helium lines, we consider this identification definitive. We opt to analyze these lines to compare with other SN subtypes and previously-observed CaSTs.

Figures 5 and 6 show the time evolution of the He I 1.083 and 2.058 μm along with the He 5876 Å line from A. P. Ravi et al. (2026). The 1 μm He feature has a P Cygni profile in all three NIR spectra of SN 2025coe, with an absorption component that persists through +34 days past peak brightness in the optical. However, the 2 μm feature shows a weaker emission component but consistently strong absorption. At later phases, the absorption profile of the 1 μm feature develops a double-peaked profile that is likely due to line blending (see Figure 5).

As seen in the left panel of Figure 5, the absorption component of the optical He feature shows a similar width and depth to the absorption component of the NIR 2 μm He line while the 1 μm He feature exhibits a broader absorption component. The resemblance between the optical He feature and 2 μm NIR He feature suggests the 1 μm feature contains contributions from neighboring lines. Neighboring features that could contribute to the blue side of this feature include Mg II λ1.0927, C I λ1.069, and S I λ1.082 (D. Milisavljevic et al. 2017; M. Shahbandeh et al. 2022).

The subsequent evolution of the line profile into a double-peaked shape resembles the NIR spectral evolution of He-rich SESNe (M. Shahbandeh et al. 2022), and therefore we conclude Mg II λ1.0927 as the more likely contributor to the He 1 μm feature. C I λ1.069 is considered as it is one of the strongest NIR C lines, but the lack of another strong C line (C I λ2.1259) near the 2 μm feature suggests C I may not be as present in the 1 μm region.

We fit Gaussians to the normalized line profiles (normalized to the peak flux in the region around the line) in order to measure the absorption velocity for both lines and the FWHM of the emission feature for the 1.083 μm line. We obtain observed flux errors from the reductions for the two Keck spectra and we calculate errors using the Fourier analysis described in Y.-Q. Liu et al. (2016) and R. Baer-Way et al. (2025a) for the MMIRS spectrum, checking that our method gives similar errors (within 20%) to those from the reductions for the Keck spectra. We fit with MCMC (D. Foreman-Mackey et al. 2013), with 10,000 chains and a 4000 step burn in in the region ±20,000 km s$^{-1}$ from the zero velocity point for the 1.083 and 2.085 μm lines. We use a one-Gaussian fit to measure properties of the absorption profile and two Gaussians for the 1 μm feature to measure the FWHM of the emission profile. We do not measure an FWHM for the 2 μm emission feature due to the relative weakness of this feature and the difficulty in defining the continuum. The absorption profile is defined relative to a continuum that is part of the fit; in other words, we allow the fit to determine the rough level of the continuum. This continuum is shown with an orange line in Figure 6. Errors are 1$\sigma$ errors taken from the posterior distribution.

The measured velocity of the NIR He absorption features decreases after peak brightness in the optical (see Figure 7). The decreasing velocities of the lines over time are consistent with the photosphere receding deeper into the ejecta and, thus the spectra revealing deeper layers of the ejecta that are at lower velocities, given the SN's homologous expansion. The velocities measured in the NIR follow the same trend as the optical He 5876 Å line, but are somewhat larger. For the 1.083 $\mu$m line, this can be explained by blending or the fact that the line is actually seen at larger radii. For the 2.058 $\mu$m line, the values are relatively similar, except at the last epoch. This could be explained by the growing influence of the neighboring Ca 1.97 $\mu$m line by the last NIR epoch, when the SN is becoming nebular.

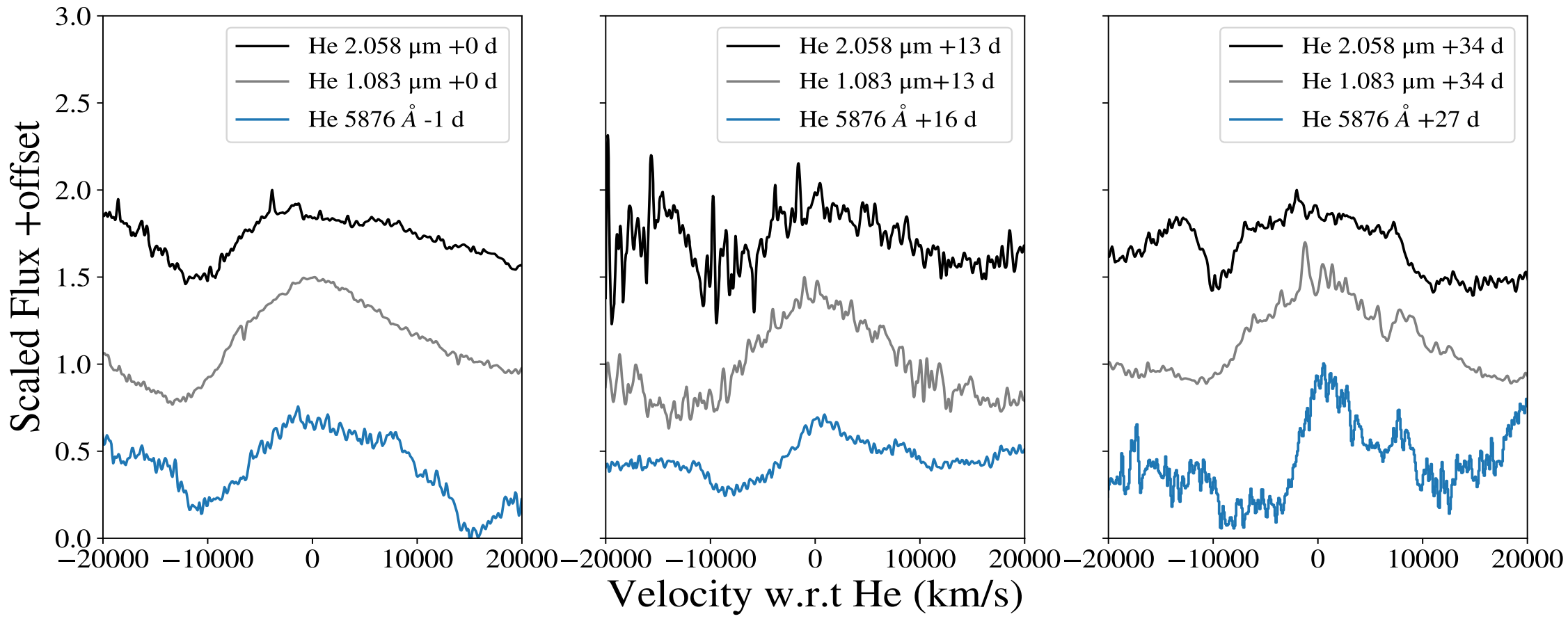

**Figure 5.** A comparison between SN 2025coe optical and NIR He line profiles at the three epochs at which NIR spectra were taken, with all phases indicated with respect to the $r$-band peak. At earlier phases, the optical and NIR He features exhibit similar P Cygni profiles. As SN 2025coe evolves over time, the NIR He line profiles begin to take on a more boxy shape that is most evident in the 2 $\mu$m He feature at +34 days.

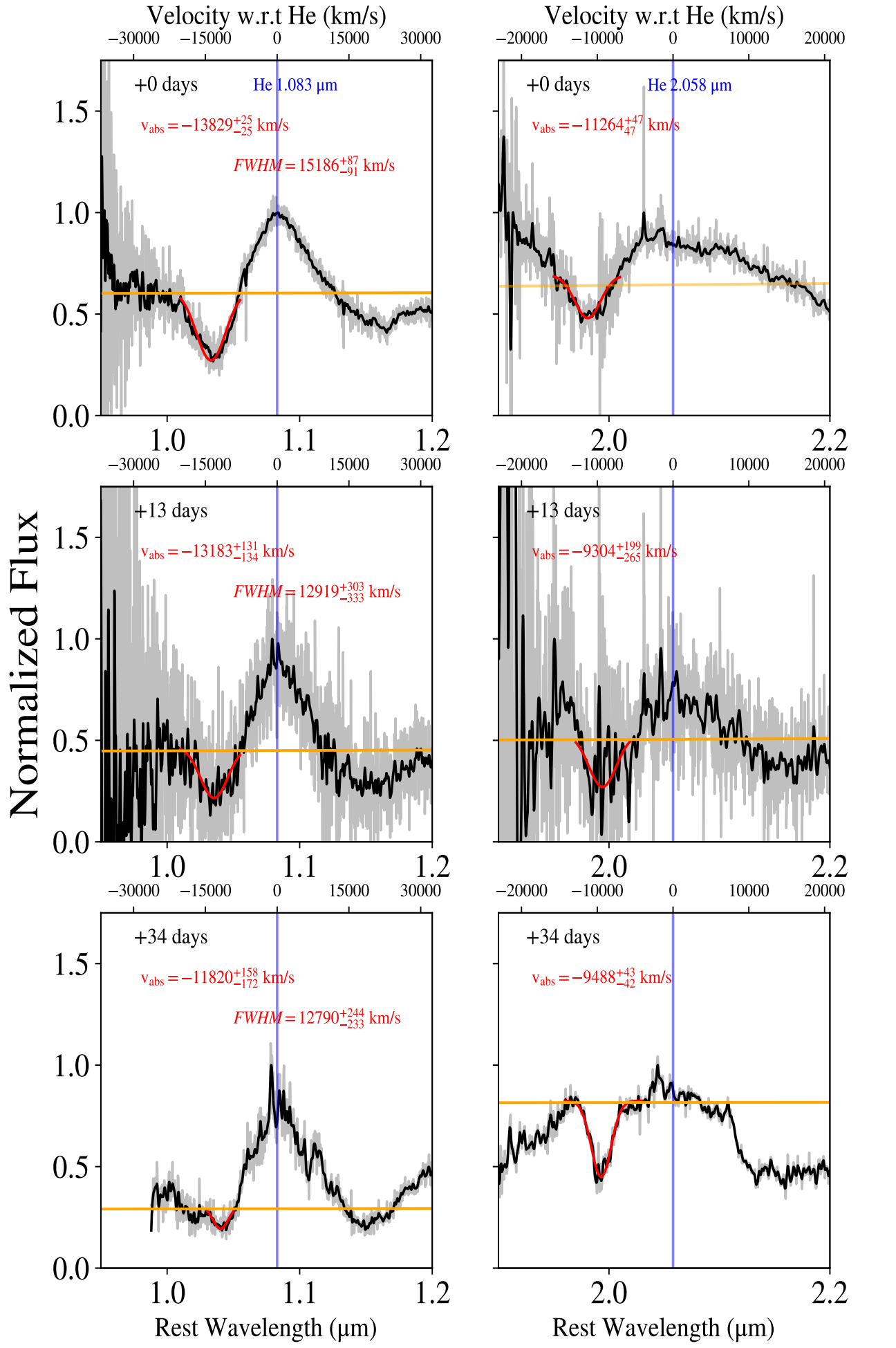

**Figure 6.** Markov Chain Monte Carlo (MCMC) fits to the He 1.083 and 2.058 $\mu$m profiles. The emission component is fit in the 1.083 $\mu$m profile but not the 2.058 $\mu$m profile due to its relative lack of strength. The fit continuum is noted with the orange line. The evolution of these NIR He features is further discussed in Section 3.2.

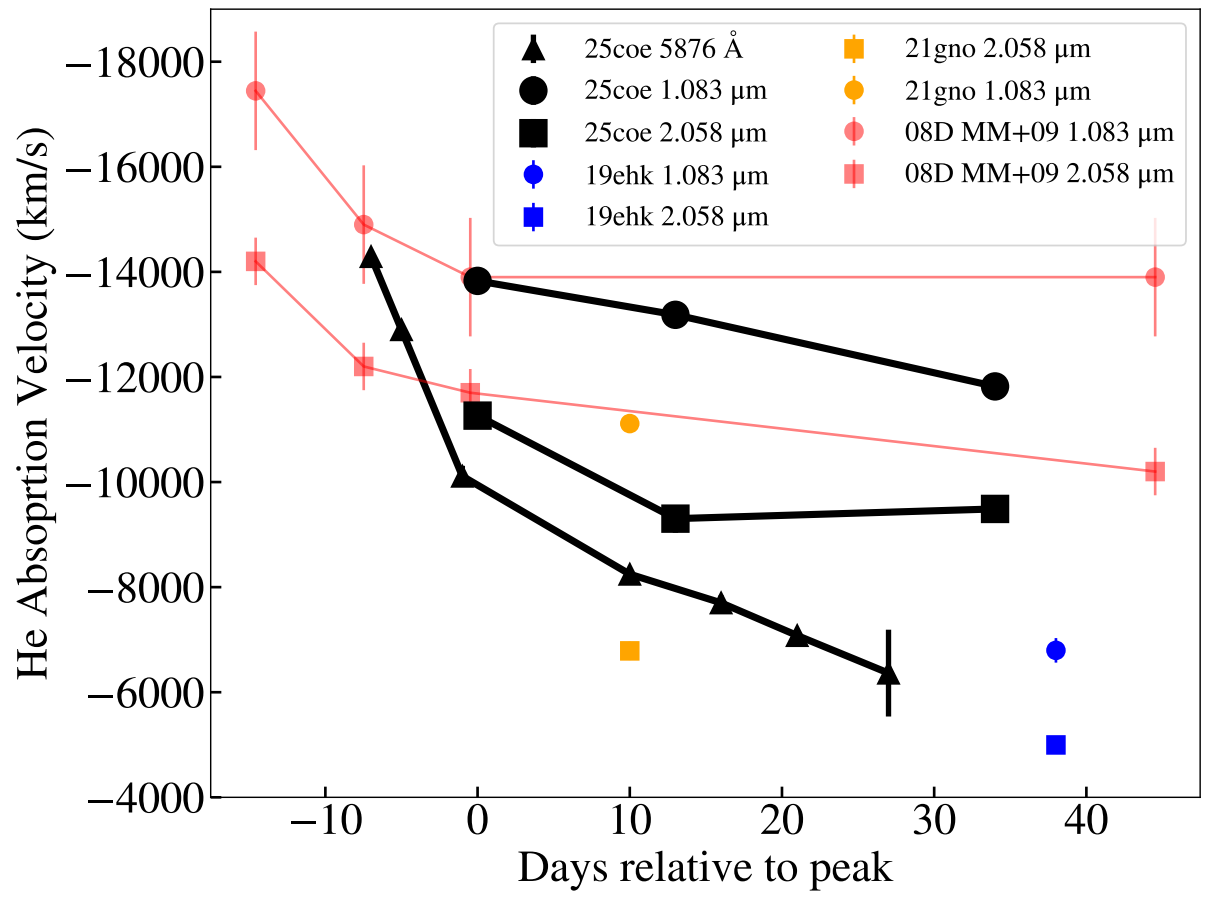

**Figure 7.** The evolution of the two NIR helium line absorption velocities in SN 2025coe. We also show the optical He 5876 Å velocity evolution (A. P. Ravi et al. 2026) for comparison. In addition, we show measurements from other CaSTs with NIR spectra, as well as from SN Ib SN 2008D (M. Modjaz et al. 2009). All dates are relative to the main (presumably nickel-powered) $r$-band peak given the double-peaked nature of some of these transients. The velocities measured for SN 2025coe are higher than those of the other two CaST-Ibs, and similar to those of SN Ib SN 2008D.

We use the same method described above to measure He NIR absorption velocities in previously published NIR spectra of SN 2021gno and SN 2019ehk (W. V. Jacobson-Galán et al. 2020a, 2022). We also plot the absorption values for SN 2008D from M. Modjaz et al. (2009), with the 1.083 $\mu$m value at −14.5 days, which was remeasured using our method to be −17,500 km s$^{-1}$. Any discrepancies with previously published values are attributed to line blending effects and possible contamination from neighboring spectral lines. All values from M. Modjaz et al. (2009) are consistent with values found using our method for the same NIR spectra of SN 2008D.

We find that the absorption velocities in SN 2025coe are generally higher than the velocities at similar epochs in SN 2019ehk and SN 2021gno, and more similar to those of SN Ib SN 2008D. The reasons for this difference are unclear—both observationally, given the single-epoch NIR datasets for the other two objects, and theoretically, given the degeneracies between ejecta mass and explosion energy. The other two CaSTs (SN 2021gno and SN 2019ehk) also show a 2 $\mu$m He feature with a consistently lower velocity than the 1 $\mu$m He feature. As mentioned, one possibility is simply that the He 2 $\mu$m feature is becoming optically thin at smaller radii (and thus has a lower velocity) due to a higher Einstein A value (W. V. Jacobson-Galán et al. 2020a). This is commonly seen in SNe Ib (M. Shahbandeh et al. 2022), and could be indicative of another observational similarity between SN 2025coe and SESNe. However, the 2 $\mu$m He I line velocity is closer to the optical He I velocities (A. P. Ravi et al. 2026), and thus we conclude that the discrepancies between the measured velocities of these two He I features is most likely due to line blending in the 1 $\mu$m feature with neighboring lines, but we cannot rule out the possibility of line formation at different optical depths.

One other point of interest is the seemingly boxy profile of the He 2.058 $\mu$m feature seen at 34 days postpeak (see Figure 5 and the lowermost right panel in Figure 6). Based on the blue and red edges of this feature, this $\sim$10,000 km s$^{-1}$ profile may be evidence for interaction with a dense shell of CSM, such as in the Type II SN (SN II) SN 2017eaw (J. Pearson et al. 2025). However, the lack of boxy profile shapes in any other spectral line in SN 2025coe, including the stronger 1 $\mu$m and optical He features at similar epochs (A. P. Ravi et al. 2026; see Figure 5), suggests some other effect is at play.

The presence of a strong absorption trough in He I 2.058 $\mu$m also implies a contribution from unshocked ejecta to the line, and the rising continuum blueward of the absorption trough (potentially due to Ca 1.97 $\mu$m) makes interpreting the line shape difficult. Furthermore, interaction with CSM created by progenitor mass loss at the level of $\dot{M} > 10^{-5}\,M_\odot$ yr$^{-1}$ at this epoch (see Section 3.4) is ruled out by the radio and X-ray observations. However, boxy profiles can be generated with even very low-density CSM (J. Pearson et al. 2025), so the X-ray and radio nondetections do not completely rule out the presence of CSM from mass loss at $<10^{-5}\,M_\odot$ yr$^{-1}$.

We note that C. Chen et al. (2026) claim to have detected a tentative third peak in the optical light curve of SN 2025coe at 43 days postexplosion and conclude this third peak is due to CSM interaction. Interestingly, the claimed third peak is at the same epoch as when this boxy profile is seen in the NIR 2 $\mu$m He line. However, this third peak in the optical light curve was not detected by A. P. Ravi et al. (2026). Given the aforementioned lack of strong evidence, we cannot firmly conclude that this is truly a boxy helium profile from interaction with a CSM shell. Detailed NIR radiative transfer simulations (as done for, e.g., SN 2014L with TARDIS; J. Lu et al. 2026) are needed for SN 2025coe to understand if this boxy line shape is due to blending, as well as what amount of He could have caused the observed NIR He absorption lines. We note that the NIR radiative transfer simulations of He presented in Figure 3 of M. Williamson et al. (2021) show a similar qualitative trend for the NIR He line profiles as seen here: the He I 2 $\mu$m line does not show the emission part of the classical P Cygni profile, while the He I 1 $\mu$m line does.

### 3.3. Comparison to Near-infrared Templates

In Figure 8, SN 2025coe is compared to NIR spectral templates of a He-rich SESN (M. Shahbandeh et al. 2022) and normal SN Ia (J. Lu et al. 2023) at similar phases. The prominent He features in SN 2025coe are well matched by the He-rich SESN template at all three epochs. The similarity of the P Cygni profiles of the 1 $\mu$m He feature points toward SN 2025coe having He-rich material in the outer layers.

The 2 $\mu$m He feature in the He-rich SESN template spectrum shows good agreement to SN 2025coe at early times. By +34 days postpeak brightness, however, SN 2025coe shows a more narrow He absorption feature than the He-rich SESN template. This difference is not reflected in the 1 $\mu$m He feature at the same phase.

The Mg I feature around 1.5 $\mu$m becomes more prominent over time in the He-rich SESN template, but is notably absent from SN 2025coe. The lack of Mg I emission around 1.5 $\mu$m in SN 2025coe could be due to higher ionization in SN 2025coe compared to the He-rich SESN template. As mentioned in Section 3.2, the profile of the 1 $\mu$m He line is likely blended with Mg II $\lambda$1.0927. Possible contributions from a nearby C I line may be blending with this Mg feature as well. Although the NIR Ca lines around 1.2 $\mu$m are present in the He-rich SESN template spectrum, they become more prominent in SN 2025coe over time.

In the NIR, SN 2025coe does not resemble a typical SN Ia. The most notable differences include the presence of strong He lines and a lack of Fe and Co lines, two iron group elements characteristic of a thermonuclear explosion of a WD. SN 2025coe is also missing the "*H*-band break," an NIR spectral feature characteristic of SNe Ia that can be used to discern the extent of the $^{56}$Ni-rich region in the ejecta (see, e.g., E. Y. Hsiao et al. 2013; C. Ashall et al. 2019). Although the interpretation of the *H*-band break in SNe Ia is often model dependent, most observational evidence suggests a stratified structure in the ejecta without large-scale mixing (C. Ashall et al. 2019). The lack of an *H*-band break and minimal evolution in spectral features between +13 and +34 days in SN 2025coe may indicate a lack of distinct layers in the SN ejecta.

Furthermore, SN 2025coe shows few similarities in the NIR with a CO + hybrid HeCO WD model from Y. Zenati et al. (2023) aside from the Ca 1.2 $\mu$m feature. In Figure 8, we compare the fca1 model from Y. Zenati et al. (2023) to SN 2025coe and note some similarity in the Ca 1.2 $\mu$m features. We note that these Y. Zenati et al. (2023) models are assuming non–local thermodynamic equilibrium. While the optical model from Y. Zenati et al. (2023) matched the nebular optical spectra well (A. P. Ravi et al. 2026), the NIR portion of the model predicts much weaker He features than observed in SN 2025coe and largely overpredicts the strengths of intermediate group elements and calcium lines at later times. Figure 9 shows all three NIR spectra of 2025coe compared to a range of SN types, and shows the most similarities to stripped-envelope SNe. Based on the comparison of the observed spectra to spectral templates and models, we conclude that SN 2025coe exhibits NIR spectral features more indicative of a core-collapse event rather than a thermonuclear explosion.

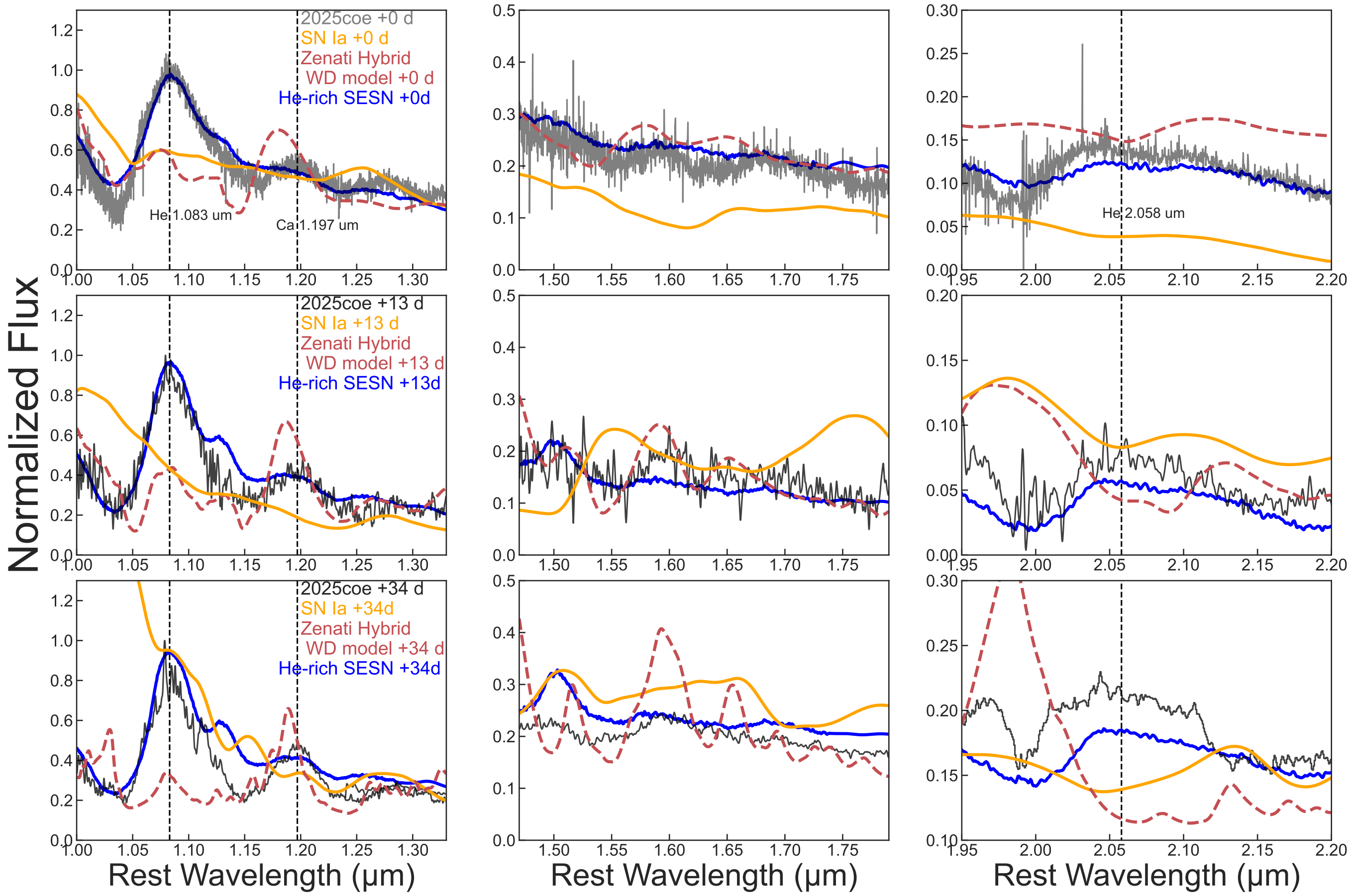

**Figure 8.** A comparison of our NIR spectra of SN 2025coe with template spectra for He-rich SESNe (M. Shahbandeh et al. 2022) and normal SNe Ia (J. Lu et al. 2023). We split the data into *J*-, *H*-, and *K*-band regions, and all epochs are relative to the optical peak (nickel-driven peak in the *B* and *V* bands). All data and models are normalized to the peak flux in the spectrum. We also plot the hybrid HeCO + CO model from Y. Zenati et al. (2023, fca1; also used in A. P. Ravi et al. 2026) as a thermonuclear model for comparison, which has been proposed as a progenitor for CaSTs. This model predicts very strong calcium lines, which are not seen in SN 2025coe. The He-rich SN Ib–like template matches the helium features well, and is a better match to the full spectrum in general.

### 3.4. Radio Nondetections and Constraints

SN 2025coe was not detected in any of the SMA, GMRT, or VLA radio observations from 20 to 153 days postexplosion. Converting flux density nondetections to luminosities, we find that SN 2025coe shows similar upper limits at similar epochs to SN 2019ehk (see Figure 4).

We use the methodology described in I. Sfaradi et al. (2025) to obtain limits on the CSM density using the late-time radio nondetections. We assume a mean shock speed for the entire evolution ($V_{\rm sh}$) such that $R = V_{\rm sh} t$. We also assume that the postshock energy density equals the energy density of the magnetic field so $B = \sqrt{9\pi \epsilon_B \rho_{\rm CSM} V_{\rm sh}^2}$ (L. DeMarchi et al. 2022), where $\epsilon_B$ is the magnetic field energy density. The electron energy density ($\epsilon_e$) and magnetic field energy density ($\epsilon_B$) are both assumed to have a value of 0.1 (R. A. Chevalier 1998), and the volume filling factor ($f$) is also assumed to have a standard value of 0.5. We also use an electron spectral index of $\gamma = 3$, which corresponds to SESN-like environments (I. Sfaradi et al. 2025). We can then use Equations (1) and (10) of R. A. Chevalier (1998) with Equation (1) of I. Sfaradi et al. (2025) to directly determine the expected $F_\nu = e^{-\tau_{\rm FFA}} F_{\nu,\rm SSA}$ for a given $\rho_{\rm CSM}$ and the assumed $V_{\rm sh}$, $\epsilon_B$, $\epsilon_e$, and $\gamma/f$ (which give us the radius ($R$) and magnetic field ($B$) of the shock). The free–free absorption (FFA) optical depth ($\tau_{\rm FFA}$) comes from Equation (3) of K. W. Weiler et al. (2002) and we assume a CSM electron temperature of $T_e = 10^5$ K at all epochs.

We then calculate at what $\rho_{\rm CSM}$ the inferred $F_\nu$ is equal to the measured upper limit for both a synchrotron self-absorption (SSA)-dominated and FFA-dominated scenario. In the SSA-dominated case, we are in an optically thin regime where $F_\nu$ is declining at the observed frequency so the method detailed above results in an upper limit on the CSM density (the CSM density must be higher than the limit as any lower value does not generate enough synchrotron power to match the upper limit on the flux density). On the other hand, the FFA-dominated scenario involves suppressed emission due to the optically thick conditions and the calculated $\rho_{\rm CSM}/\dot{M}$ is a lower limit (in other words, an $\dot{M}$ higher than our inferred limit in this case would also be consistent with the flux density upper limit as it would suppress the emission even more).

If the SMA 240 GHz nondetection at 20 days postexplosion is due to strong FFA, this would imply a mass-loss rate $> 10^{-2}\, M_\odot\, {\rm yr}^{-1}$ based on the methodology described above. At other observed radio epochs, the implied FFA limits differ significantly due to the changing frequencies of the radio data points (since $\tau_{\rm FFA}$ is frequency dependent). The GMRT nondetections at 87 days postexplosion only constrain $\dot{M} > 10^{-4}\, M_\odot\, {\rm yr}^{-1}$, while the VLA nondetections at 120–153 days suggest $\dot{M} > 10^{-2}\, M_\odot\, {\rm yr}^{-1}$. Therefore, high-density CSM would have to be present until at least 153 days after the explosion.

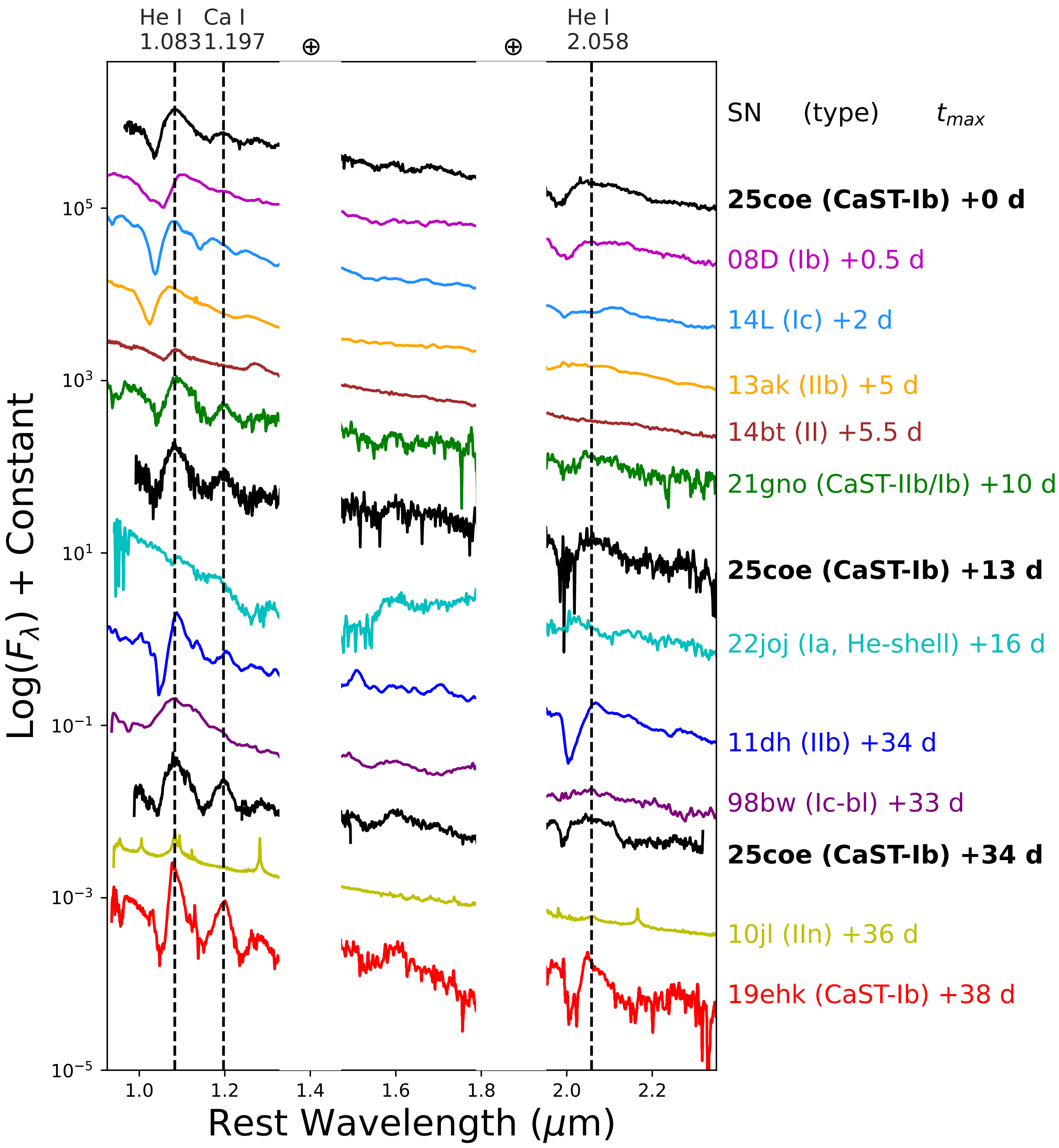

**Figure 9.** A comparison of NIR spectra of almost all known SN subtypes with our spectra of SN 2025coe shown in black. SN 2025coe shows clear strong He features similar to SNe Ib and IIb. At similar phases (with respect to peak), CaSTs like SN 2025coe exhibit more spectral features as they become nebular faster than CCSNe like SNe II and Type IIn SNe, which should originate from more massive stars. When compared to a more energetic explosion such as a broad-line Type Ic (Ic-BL), the spectral features are distinctly broader than what is observed in all three CaSTs included in this figure. One of the most notable differences is when compared to SN 2022joj, an SN Ia with early He that shows little resemblance to SN 2025coe at similar phases. Sources for the previously published spectra used for comparison are M. Modjaz et al. (2009), M. Shahbandeh et al. (2022), S. Davis et al. (2019), W. V. Jacobson-Galán et al. (2022), E. Padilla Gonzalez et al. (2024), G. H. Marion et al. (2014), F. Patat et al. (2001), H. J. Borish et al. (2015), and W. V. Jacobson-Galán et al. (2020a).

Given that there is no evidence for extended dense CSM from the optical light curves or spectra (A. P. Ravi et al. 2026), it seems highly unlikely that the radio nondetections are caused by strong absorption from dense material. Additionally, interaction with a CSM formed from a mass-loss rate this high would be expected to generate a reprocessed optical luminosity $> 5 \times 10^{41}$ erg s$^{-1}$ (N. N. Chugai 1991) for a shock speed $> 5000$ km s$^{-1}$, a reprocessing efficiency to optical luminosity of $\epsilon = 0.1$, and a CSM speed of 1000 km s$^{-1}$, which exceeds the entire bolometric luminosity estimates

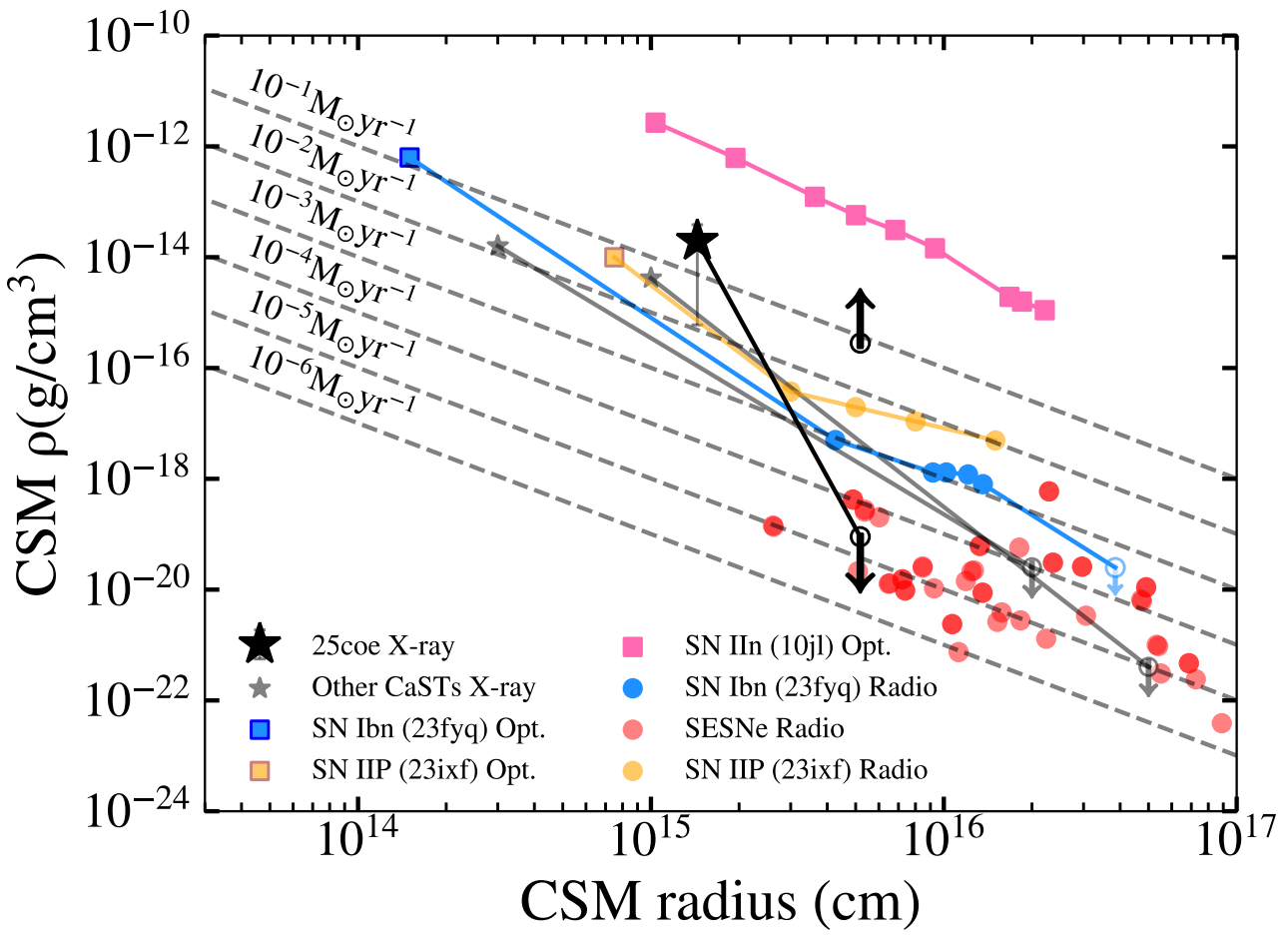

**Figure 10.** A view of the measured CSM densities for SN 2025coe, other CaSTs, and other SN subtypes in general. We add the SSA-dominated (downward-facing arrow) and FFA-dominated (upward-facing arrow) limits on the CSM density from the radio analysis, but note that the low-density SSA scenario is highly favored based also on the optical analysis from A. P. Ravi et al. (2026). All density values and upper limits are obtained directly from either optical spectroscopy or the radio and X-ray data. Optical data are denoted with squares, while radio data are denoted with circles and X-ray data are denoted with stars. All limits are denoted with arrows originating from open circles (to denote the radio limit). Lines of constant mass loss are plotted for a 500 km s$^{-1}$ wind speed. We note that we have compositions from references for all other SNe, and assumed solar composition for SN 2025coe. Data for other SNe are from T. Zhang et al. (2012), W. V. Jacobson-Galán et al. (2020a, 2022, 2023), A. J. Nayana et al. (2025), I. Sfaradi et al. (2025), R. Baer-Way et al. (2025b), and Y. Dong et al. (2024).

(at $\sim 10^{41}$ erg s$^{-1}$) past 20 days postexplosion (A. P. Ravi et al. 2026).

We thus use the SMA nondetection to constrain the CSM density in the SSA limit at 20 days postexplosion. Using the nondetection at 240 GHz at this epoch (so when the CSM would be at a radius $\sim 5.0 \times 10^{15}$ cm assuming a shock speed of 30,000 km s$^{-1}$) and we find $\dot{M} < 2.5 \times 10^{-5}\,M_\odot$ yr$^{-1}$ for a 1000 km s$^{-1}$ wind speed (typical for an SESN; K. W. Weiler et al. 2002). Whether the CSM is helium or hydrogen rich (or rich in any other element) would not make enough of a difference (<a factor of 3) to explain the nondetections in the radio if there were still dense CSM present. There is thus no evidence for high-density CSM beyond the first 10 days postexplosion. Given this first nondetection at 20 days, we find that the CSM could conservatively extend to $5.0 \times 10^{15}$ cm. We show the CSM density measured from the X-ray analysis along with this upper limit in Figure 10.

### 3.5. Multiwavelength Picture of SN 2025coe

Combining the insights from our X-ray, NIR, and radio observations, we can paint a clear picture of SN 2025coe. The X-ray and radio observations suggest nearby $((\sim 3.5 \pm 1.5) \times 10^{15}$ cm) high-density CSM, which the SN ejecta overrun (see Figure 10). The NIR observations suggest He ejecta at speeds typical for SESNe, and at higher velocities than in the previously observed CaSTs SN 2019ehk and SN 2021gno (see Figure 7). If all three nearby CaSTs had the same explosion energy, the higher velocities in SN 2025coe could indicate a lower ejecta mass than for SN 2019ehk and SN 2021gno. However, the estimated ejecta mass in SN 2025coe (A. P. Ravi et al. 2026) of 0.4–0.5 $M_\odot$ is only slightly lower than the estimates of 0.6 and 0.7 $M_\odot$ for SN 2021gno and SN2019ehk, respectively (W. V. Jacobson-Galán et al. 2020a, 2022), and not enough to account for the discrepancy in velocities of a factor ∼ 1.5–2.

We conclude that our NIR, X-ray, and radio observations of SN 2025coe are consistent with a stripped-envelope CCSN that had a very compact CSM. However, the fact that the implied mass-loss rate is so high and the CSM is so nearby makes determining details of the massive star progenitor system difficult. As mentioned, no prior SESN has ever shown such radially limited CSM at such a high mass-loss rate (see the compilation in D. Brethauer et al. 2022; Figure 10). A low-mass massive stripped star progenitor that experienced intensive mass loss before explosion would seemingly be more likely to fall into the Type Ibn category (see, e.g., K. Maeda & T. J. Moriya 2022; R. Baer-Way et al. 2025b), but the He lines in SN 2025coe are not in emission, the hallmark of SNe Ibn. We speculate that these CaSTs may represent a transitional case where the mass loss is slightly less intense and more limited in timescale than in SNe Ibn and the total ejecta and nickel mass is still lower compared to SNe Ibn. Since narrow He emission lines in SNe Ibn are thought to have been caused by the SN ejecta running into nearby CSM, SN Ibn progenitors likely lose much of their He layer while CaST progenitors still have fairly intact helium layers.

In the thermonuclear case, the CSM could have been created from pollution by a recurrent helium nova AM CVn system around a binary HeCO + CO WD system as suggested in C.-G. Touchard-Paxton et al. (2025). This model could explain the observed CSM more naturally, but as seen from our NIR spectra, models of such systems would need to reproduce the observed strong He P Cygni NIR lines of these CaSTs, including SN 2025coe, which are very similar (in strength and, for SN 2025coe, in high velocity) to those of normal He-rich stripped-envelope CCSNe, where the He is part of the ejecta and not of the CSM.

As one final point of emphasis: to date, no thermonuclear SN has been conclusively detected at X-ray wavelengths. If the class of CaSTs is indeed of thermonuclear origin, then SN 2025coe, together with SN 2019ehk and SN 2021gno, would constitute the first thermonuclear SNe of any subclass to exhibit X-ray emission—a result of major significance for both explosion physics and progenitor models. Conversely, the presence of luminous X-ray emission in these CaSTs may instead indicate that the X-ray-bright members of this class arise from core-collapse explosions, while genuinely thermonuclear CaSTs are intrinsically X-ray faint. Indeed, for the thermonuclear CaST-Ia SN 2016hnk, no X-ray emission was detected (P. H. Sell et al. 2018) to deep limits at ∼30 days after explosion (see Figure 2).

## 4. Conclusion

We present detailed panchromatic X-ray, NIR and radio observations of the CaST SN 2025coe from 2 to 153 days postexplosion and thoroughly compare them to those of other CaSTs, to normal He-rich stripped-envelope CCSNe, to thermonuclear explosions, and some models. Our main results are as follows.

1. SN 2025coe produced luminous X-ray emission at $\sim 3 \times 10^{40}$ erg s$^{-1}$ for the first 8 days postexplosion. From modeling the emission as the product of ejecta–CSM interaction, we infer that the SN ejecta were interacting with $0.12 \pm 0.11\,M_\odot$ ($\rho_{\rm CSM} = (1.96 \pm 1.90) \times 10^{-14}\,{\rm g\,cm^{-3}}$) of CSM, which extends to at least $\sim 2 \times 10^{15}$ cm.
2. Subsequent radio and X-ray nondetections suggest a lack of dense material beyond $5 \times 10^{15}$ cm. The combined X-ray and radio data imply extremely dense CSM (presumably caused by intensive mass loss in the final years preexplosion) out to $(3.5 \pm 1.5) \times 10^{15}$ cm, which has not been seen in any normal SESNe, if SN 2025coe is a stripped core-collapse event.
3. Our multiple NIR spectra of SN 2025coe represent the first NIR spectral time series of a CaST. They show very strong helium P Cygni features at 1.083 and 2.058 $\mu$m from 10 to 44 days postexplosion. These NIR He lines are consistent with those seen in other CaSTs of Ca-Ib and Ca-IIb subtypes and in the optical spectra of SN 2025coe in the companion paper by A. P. Ravi et al. (2026). In general, the whole shape of the NIR spectra of SN 2025coe across 1–2.3 $\mu$m is markedly similar to the NIR template spectra of normal He-rich SESNe (M. Shahbandeh et al. 2022) and very different from those of normal thermonuclear SNe Ia (J. Lu et al. 2023) and from the thermonuclear merger model of Y. Zenati et al. (2023). The strengths of the He features are consistent with those seen in NIR spectra of normal helium-rich SESNe, but are much stronger than those simulated by Y. Zenati et al. (2023) for hybrid WDs with helium shell detonations. Conversely, the observed calcium features in SN 2025coe are not as strong when compared with these hybrid WD models.
4. The fact that all three of the nearest CaSTs of subtypes Ca-Ib and Ca-IIb (SN 2019ehk, SN 2021gno, and SN 2025coe) have exhibited X-ray emission due to interaction with nearby dense CSM suggests this is likely a common feature of the subclass, and needs to be accounted for when explaining their progenitor systems. The presence of full He P Cygni profiles in all three nearby objects also indicates a large amount of He in the ejecta, similar to those of He-rich SNe Ib, at absorption velocities ranging from 6,000 km s$^{-1}$ (for SN 2019ehk and SN 2021gno) to 10,000–12,000 km s$^{-1}$ (for SN 2025coe).

Our conclusions are consistent with those from the optical dataset of SN 2025coe (A. P. Ravi et al. 2026), particularly in terms of the better template matches to SESNe as opposed to thermonuclear models. However, the presence of CSM is not clearly seen in the optical spectra, perhaps due to geometric effects. The overall X-ray, NIR, and radio results, and in particular the strong He absorption lines seen in the NIR, are consistent with a core-collapse origin for this event.

Nonetheless, massive star progenitor system models need to explain why the CSM is far denser than seen in normal SESNe and extends only to smaller radii, in addition to why the oxygen lines are so much weaker than the Ca lines. Additionally, the large offset of SN 2025coe from the center of the host galaxy (A. P. Ravi et al. 2026) may be difficult to reconcile with a core-collapse origin. Future radio and X-ray observations closer to the explosion date (within 10 days postexplosion, see Figures 2 and 4), and with higher sensitivity (i.e., with Chandra or for longer exposures with Swift/XRT) will be vital to first detect and then constrain the CSM of these objects in detail, in order to compare with progenitor system simulations.

Furthermore, future NIR spectral synthesis calculations with the inclusion of nonthermal excitation of He are needed to estimate the amount of He in the ejecta of SN 2025coe and the two other CaSTS, which are of subtypes Ca-Ib and Ca-IIb. These estimates could help constrain the progenitor models, both for CCSNe and WDs with He detonation models. For models involving a WD with He, such as the promising one in C.-G. Touchard-Paxton et al. (2025, which is different from Y. Zenati et al. 2023, whose radiative transfer calculations we show in Figure 8), where the CSM could have been polluted by a recurrent helium nova AM CVn system before it became a binary HeCO + CO WD system (that explodes via helium detonation), NIR radiative transfer predictions are needed to compare with the observed NIR spectra of the three nearest CaST systems. Similarly, while the simulations of double-detonation models of sub-Chandrasekhar-mass CO WDs with He shells from A. Polin et al. (2021) seem to fit the nebular spectra of SN 2025coe as shown in A. P. Ravi et al. (2026), NIR spectral simulations of their model are needed to compare to our NIR data in order to test that particular thermonuclear model.

As the number of known CaSTs grows over time, their faint and rapidly evolving natures will require prompt and well-coordinated follow-up observations to reveal more information on their enigmatic origins. Since many CaSTs, including SN 2025coe, are initially classified as SESNe based on optical spectroscopy, identifying key observables in other wavelength regimes can help properly identify CaSTs at earlier phases. Early X-ray emission has now been detected in three nearby CaSTs and is at early enough phases to trigger subsequent observations at other wavelengths. The new observations of SN 2025coe presented in this work provide key impetus for continued future NIR observations of CaSTs, especially at later times with JWST and the Nancy Grace Roman Space Telescope. Finally, late-time radio observations provide physical constraints on the extent of the CSM, which are necessary to distinguish different physical scenarios. SN 2025coe further emphasizes how multiwavelength observations provide the best opportunity for studying the progenitors and local environments of CaSTs.

## Acknowledgments

The authors thank Wolfgang Kerzendorf, Jing Lu, and Wynn Jacobson-Galán for the helpful discussion. M.M. and the METAL group at UVA acknowledge support in part from ADAP program grant No. 80NSSC22K0486, from the NSF grant AST-2206657, from the National Science Foundation under Cooperative Agreement 2421782, and the Simons Foundation grant MPS-AI-00010515 awarded to the NSF-Simons AI Institute for Cosmic Origins—CosmicAI, at https://www.cosmicai.org/. R.B.W. is supported by the National Science Foundation Graduate Research Fellowship Program under grant No. 2234693 and acknowledges support from the Virginia Space grant Consortium. S.V. and A.P.R. acknowledge support by NSF grants AST-2407565. P.C. acknowledges the support of this work provided by the National Aeronautics and Space Administration through Chandra Award Numbers GO3-24056X, DD3-24141X, and GO4-25044X. Time-domain research by the University of Arizona team and D.J.S. is supported by National Science

Foundation (NSF) grants 2308181, 2407566, and 2432036. The Very Large Array is operated by the National Radio Astronomy Observatory, a facility of the U.S. National Science Foundation (NSF) operated under cooperative agreement by Associated Universities, Inc. The GMRT is run by the National Centre for Radio Astrophysics of the Tata Institute of Fundamental Research. Some observations reported here were obtained at the MMT Observatory, a joint facility of the University of Arizona and the Smithsonian Institution. Some of the data presented herein were obtained at the W. M. Keck Observatory, which is operated as a scientific partnership among the California Institute of Technology, the University of California, and NASA. The Observatory was made possible by the generous financial support of the W. M. Keck Foundation. The authors wish to recognize and acknowledge the very significant cultural role and reverence that the summit of Maunakea has always had within the indigenous Hawaiian community. We are most fortunate to have the opportunity to conduct observations from this mountain.

## ORCID iDs

Sahana Kumar https://orcid.org/0000-0001-8367-7591
Raphael Baer-Way https://orcid.org/0009-0004-7268-7283
Aravind P. Ravi https://orcid.org/0000-0002-7352-7845
Maryam Modjaz https://orcid.org/0000-0001-7132-0333
Poonam Chandra https://orcid.org/0000-0002-0844-6563
Stefano Valenti https://orcid.org/0000-0001-8818-0795
Lindsey A. Kwok https://orcid.org/0000-0003-3108-1328
Samaporn Tinyanont https://orcid.org/0000-0002-1481-4676
Ryan J. Foley https://orcid.org/0000-0002-2445-5275
D. Andrew Howell https://orcid.org/0000-0003-4253-656X
Daichi Hiramatsu https://orcid.org/0000-0002-1125-9187
Jennifer E. Andrews https://orcid.org/0000-0003-0123-0062
K. Azalee Bostroem https://orcid.org/0000-0002-4924-444X
Collin Christy https://orcid.org/0000-0003-0528-202X
Noah Franz https://orcid.org/0000-0003-4537-3575
Brian Hsu https://orcid.org/0000-0002-9454-1742
Jeniveve Pearson https://orcid.org/0000-0002-0744-0047
David J. Sand https://orcid.org/0000-0003-4102-380X
Manisha Shrestha https://orcid.org/0000-0002-4022-1874
Nathan Smith https://orcid.org/0000-0001-5510-2424
Bhagya Subrayan https://orcid.org/0000-0001-8073-8731